\documentclass[10pt, sigconf, nonacm]{acmart}
\usepackage{array}
\usepackage{multirow}
\usepackage{graphicx}
\usepackage{algorithm}
\usepackage{algorithmicx}
\usepackage{algpseudocode}
\usepackage{amsmath}
\usepackage{amssymb}

\AtBeginDocument{%
	\providecommand\BibTeX{{%
			\normalfont B\kern-0.5em{\scshape i\kern-0.25em b}\kern-0.8em\TeX}}}

\begin{document}

\title{MIN: Co-Governing Multi-Identifier Network Architecture and Its Prototype on Operator's Network}




\begin{teaserfigure}
	\includegraphics[width=\textwidth]{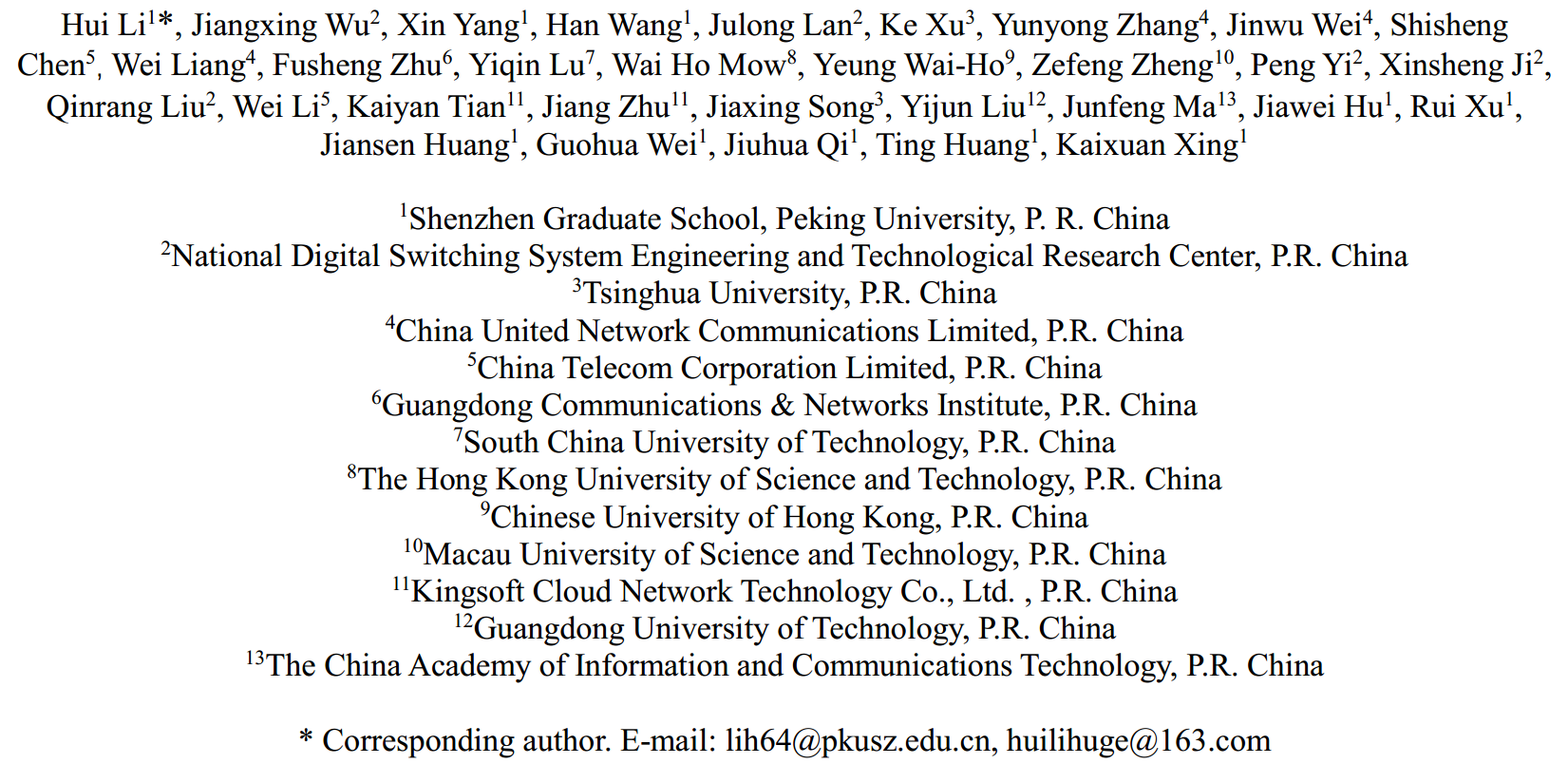}
	\label{fig:teaser}
\end{teaserfigure}

\begin{abstract}

IP protocol is the core of TCP/IP network layer. However, since IP address and its Domain Name are allocated and managed by a single agency, there are risks of centralization. The semantic overload of IP address also reduces its scalability and mobility, which further hinders the security.

This paper proposes a co-governing Multi-Identifier Network (MIN) architecture that constructs a network layer with parallel coexistence of multiple identifiers, including identity, content, geographic information, and IP address. On the management plane, we develop an efficient management system using consortium blockchain with voting consensus, so the network can simultaneously manage and support by hundreds or thousands of nodes with high throughput. On the data plane, we propose an algorithm merging hash table and prefix tree (HTP) for FIB, which avoids the false-negative error and can inter-translate different identifiers with tens of billions of entries. Further, we propose a scheme to transport IP packets using CCN as a tunnel for supporting progressive deployment. We deployed the prototype of MIN to the largest operators' network in Mainland China, Hongkong and Macao, and demonstrated that the network can register identifier under co-governing consensus algorithm, support VoD service very well.

\end{abstract}


\keywords{Network Architecture, Consortium Blockchain, HTP-FIB, APoV, MIN, MIS, MIR}


\maketitle

\section{Introduction}

The Internet architecture based on IP has continuously evolved during the last decades. Its end-to-end transmission mode accelerated its deployment worldwide. However, in IP based network, the allocation and management of IP addresses are controlled by a single agency. Such centralization brings risks. On the other hand, IP addresses represent both the location and the identity information of nodes, causing semantic overload. These problems hinder the scalability of routing and its adaptation to mobile devices. The lack of identity tracking and managing also reduce the security of IP-based networks \cite{J2, J3, J4}. The defects in security and tractability make the IP network unsuitable for supporting emerging scenarios, such as Content-Centered Network (CCN), high-speed mobile network, Internet of Things, and Industrial Internet. The situation poses urgent demand for a new generation of future network architectures.

To improve the performance of IP-based networks, European Union launched the first phase of Future Internet Research and Experimentation (FIRE) program \cite{J7} to explore future network and service mechanisms. The project focused on exploring the self-adapting management mechanism in the future network to increase its level of intelligence. However, this study did not produce a new network architecture design. The NDN project \cite{J5} backed by the National Science Foundation (NSF) aims to develop a network architecture based on content. The MobilityFirst project \cite{J6} prioritizes the identity of nodes and efficiently manage the movement of nodes. It uses the General Delay-Tolerant Network (GDTN) to enhance the robustness and availability of network.

To decentralize the management of the network architecture, blockchain \cite{J10} and other solutions \cite{J11, J12, J13, J14, J15} have recently been applied to build a future network under co-governing. Namecoin \cite{J16} and Blockstack \cite{J17} first applied blockchain to decentralize the management of domain name system. However, its underlying system based on public blockchain creates bottleneck for its performance. To solve the problem, Benshoof et al. proposed an alternative solution of DNS system based on blockchain and distributed hash table named $ D^{3}NS $ \cite{J19}, which provides solutions to current DNS vulnerabilities such as DDoS attacks. However, it risks leaking users' IP information and increases the difficulty to deploy in large-scale. To mitigate the problem, HyperPubSub system \cite{J20} uses the passive publish/subscribe receiving mode to reduce the traffic load and the delay caused by blockchain.

The above methods improve performance of network and level of decentralization, respectively, but are unable to meet both requirements\cite{K5} simultaneously. This paper proposes a Multi-Identifier Network (MIN)  architecture that constructs a network layer with parallel coexistence of multiple identifiers, including identity, content, geographic information, and IP address. To solve the two major defects of the traditional network, we decentralize the management of post-IP identifier using consortium blockchain, address or inter-translate identifiers with tens of billions of entries using HPT for FIB. We also enhance the progressive deployment through novel IP-CCN-IP tunnel scheme. As part of the work, we implement the generation, management, and resolution in MIN and ensure it compatible with the traditional IP network, to support progressive deployment.

We experimented and verified MIN in operators' network deploy from Beijing to Guangzhou, Shenzhen, Hongkong and Macao for streaming high-definition video. We also tested the transmission including IP to CCN to IP, IP to CCN, CCN to IP, and CCN to IP to CCN. The results demonstrate that the network achieves excellent performance and can support real-world applications with further development.

As contributions, this paper proposes a novel Multi-Identifier Network. Specifically,

1) On the management plane, we design an efficient consensus mechanism of consortium blockchain to achieve decentralized management of identifiers, which is the multi-identifier system(MIS). MIS takes the same role as DNS in IP, actually the function of DNS is only a subset of MIS.

2) On the data plane, we improve HPT-FIB by combining hash table and prefix tree, which is integrated into the multi-identifier router (MIR) of MIN as a key part. Such improvement enables identity-centered forwarding and inter-translation between different kind of identifiers in a magnitude of tens of billions of entries.

3) We also introduce a scheme to transport IP packets using CCN as a tunnel to support progressive deployment of the data plane (being compatible with IP-based network). We implemented various transmission scenarios, such as IP-CCN-IP, IP-CCN, CCN-IP, and CCN-IP-CCN.

The rest of this paper is organized as follows. Section 2 describes the proposed Multi-Identifier Network. Section 3 demonstrates the key technologies. Section 4 presents the quantitative results of the prototype, and section 5 provides some concluding remarks and discusses ongoing and future research directions.

\section{Multi-Identifier Network}

For the co-governing Multi-Identifier Network, its decentralized management and large resolution capability enable a progressive transition from the existing network architecture to a new one.

\subsection{Network Architecture}

\begin{figure}[h]
\setlength{\abovecaptionskip}{0.cm}
\setlength{\belowcaptionskip}{-0.cm}
	\centerline{\includegraphics[width=3.3in]{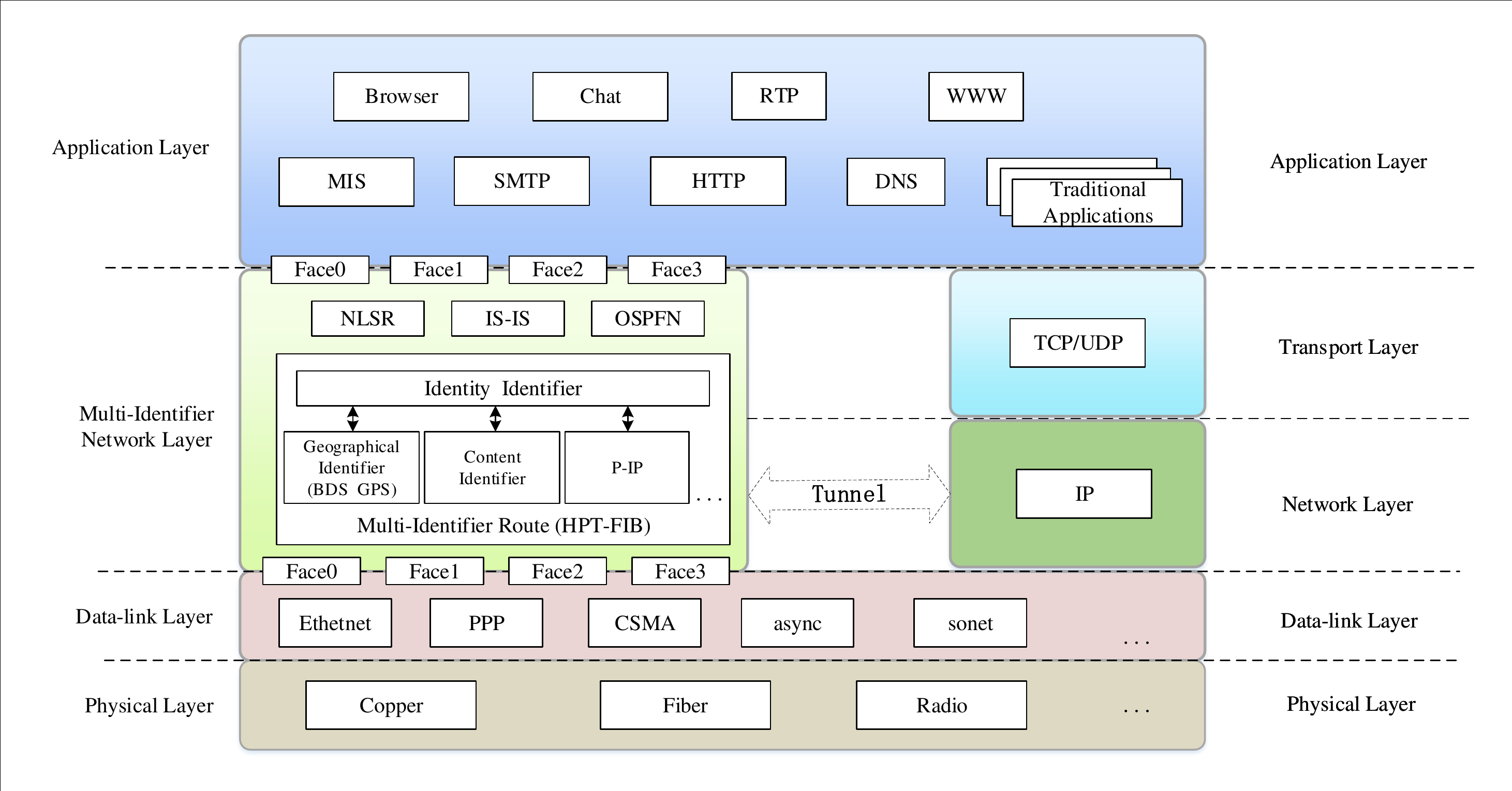}}
	\caption{Protocol Architecture of MIN}
	\label{min}
\end{figure}

MIN supports the coexistence of Network Identifiers including identity, content, geographic information, and IP address. Identifiers in the network are identity-centric. For example, the content identifiers of all resources are bound to the identity identifiers of their publishers. The protocol architecture of MIN is shown in Figure.\ref{min}.

\begin{figure}[h]
\setlength{\abovecaptionskip}{0.cm}
\setlength{\belowcaptionskip}{-0.cm}
	\centerline{\includegraphics[width=3.3in]{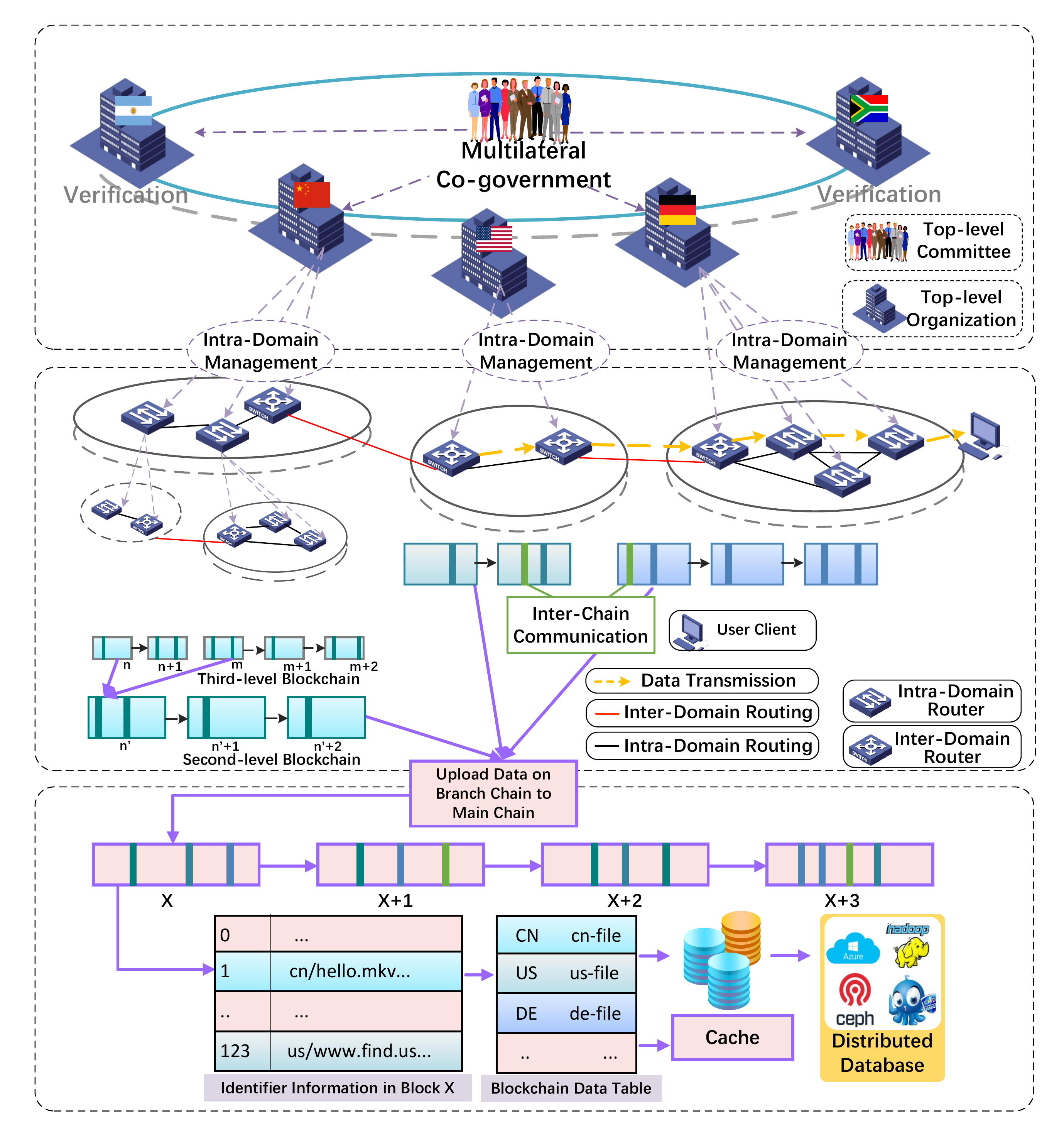}}
	\caption{Network hierarchy of MIN}
	\label{fig1}
\end{figure}

The network hierarchy of MIN is shown in Figure.~\ref{fig1}. It divides the whole network into hierarchical domains from top to bottom. The nodes in the top-level domain belong to the organizations of the major countries which jointly maintain a consortium blockchain. The respective regional organizations govern the other domains. Among them, the registration and management mode of identifiers and the specific implementation details can vary. This low coupling guarantees the security of the network and enables customization of each domain \cite{K8} \cite{K7}.

The function of a complete node in the network is to participate in the intra-domain management of users and the registration process of identifiers on the blockchain, as well as provide inter-translation and resolution services (in this case called Multi-Identifier Router, MIR). Also, there are supervisory nodes, individual users, and enterprise users. Supervisory nodes are set up as the data access interfaces between the upper and lower domains. Each supervisory node has multiple identifiers.

The architecture of MIN includes a management plane and a data plane. The management plane is responsible for generation and management of identifiers. After the supervisory nodes verify the identifier and reach a consensus through the consensus algorithm, it records the relevant attribution information and operation information on the blockchain, to make the data in the whole network unified, tamper-resistant and traceable. The reason for storing only the important data on-chain is to ensure efficiency, while all the information of the identifiers are stored off-chain. The data plane provides the resolution for identity, content, geographic information and other identifiers, and is also responsible for packet forwarding and filtering.

\subsection{Operation Flow}

In MIN, all resources are required to register an identifier with a regulatory organization within the domain. Devices can only access a resource in the network after its identifier has been approved by most organizations and successfully written on the blockchain. The registration process is as follows.

S1: The user who owns the resource submits a request for identifier registration to the node of a regulatory organization.

S2: After receiving the user's request, MIR transmits the registration data to its corresponding domain according to a specific routing protocol.

S3: The blockchain node of the corresponding domain reviews the compliance of the resource after receiving its identifier registration request. If so, the resource's identifier is then voted by all the blockchain nodes in the domain to reach consensus.

S4: The blockchain node then returns the registration result to the original requesting node. Since the complete identifier information is stored in the off-chain database rather than the on-chain block, all databases are synchronized frequently throughout the network to ensure consistency.

MIR's process to resolve the identifier is as follows.

S1: MIR judges that the identifier is (1) IP address, then query in HPT-FIB. If it exists, it will be resolved. Otherwise, access the traditional IP network through proxies; (2) identity, content and other identifiers, then query in the cache and HPT-FIB. If it exists, it will be resolved. Otherwise, go to S2.

S2: If MIR cannot find the identifier, recursively query the upper domain until acquiring it.

S3: If the identifier is not found up to the top-level domain, then query the lower domain according to the information carried by the identifier until the lowest. If it exists, MIR will return the resolved result. Otherwise, return an error message.

In MIN, users' behaviors of publishing and accessing are protected and managed, and the blockchain undeniably records illegal actions. Therefore, MIN will keep the cyberspace in an orderly and secure state, which will direct traffic to the post-IP multi-identifier network tied to the user's identity.

\section{Key Technologies}

As mentioned above, we use consortium blockchain, HPT for FIB, and tunnel algorithm to improve the security, performance, and scalability. In this section, we introduce those technologies.

\subsection{Consensus Algorithm for Consortium Blockchain}

 On the management plane of the MIN architecture, we come up with the APoV (Advanced Proof of Vote) consensus based on our self-designed PoV (Proof of Vote) \cite{K10}, a non-forking consensus algorithm for consortium blockchain. The core lies in the separation of voting rights and bookkeeping rights. The bookkeeping nodes work in a joint effort to conduct decentralized arbitration according to the votes of the consortium nodes.

~\\
\textbf{(1)Concept Description}

APoV defines that data on the blockchain is stored in block groups. A block group consists of a block group header and a block group body. Each block group header contains the height and voting result. The block group body includes blocks approved by the majority of consortium nodes. Where, each block consists of the hash value of the previous block group, the Merkle root, the public key of the bookkeeping node, the timestamp, and the set of transactions.

The APoV consensus divides the blockchain nodes into three identities: consortium node, bookkeeping node, and leader node.

\begin{itemize}
	\item The consortium node is responsible for voting on the generated blocks and potential bookkeeping nodes. The number of consortium nodes in each round is a fixed constant, denoted as $ n_c $. Based on the principle of ``the minority is subordinate to the majority", the voting results are regarded as proof of the validity of the block and the identity of the bookkeeping node.
	\item The bookkeeping node is responsible for generating blocks in the current consensus round. The number of bookkeeping nodes is $ n_b $. At the end of the term, the consortium nodes vote on the potential bookkeeping nodes to produce the next bookkeeping nodes.
	\item The leader node is responsible for counting votes and writing the voting result into the block group header as proof. Each consensus round has a different leader node, whose number is recorded in the previous block group header.
	
\end{itemize}

There are two types of voting messages in APoV for the transactions of identifiers and election: confidence vote and verification vote.

\begin{itemize}
	\item The validation vote is a validation of the block group. The consortium node votes for the blocks they agree to generate. Each block must obtain more than half of the votes to be considered as a legal block. Similarly, to correct a result, more than half of the consortium nodes must agree.
	\item The confidence vote is a successful proof for the next bookkeeping nodes. Before the end of the current bookkeeping nodes' term, each potential bookkeeping node proposes a transaction of election and receives votes from consortium nodes. The voting result indicates the trust of consortium nodes in these nodes competing for bookkeeping rights, thus can also be considered as the reliability of them. The nodes with higher reliability ranking are deemed successful in the election, with bookkeeping rights from the next consensus round until the end of their term.
	~\\	
\end{itemize}

\textbf{(2)Process of Consensus}

\begin{figure}[h]
\setlength{\abovecaptionskip}{0.cm}
\setlength{\belowcaptionskip}{-0.cm}
	\centerline{\includegraphics[width=3.3in]{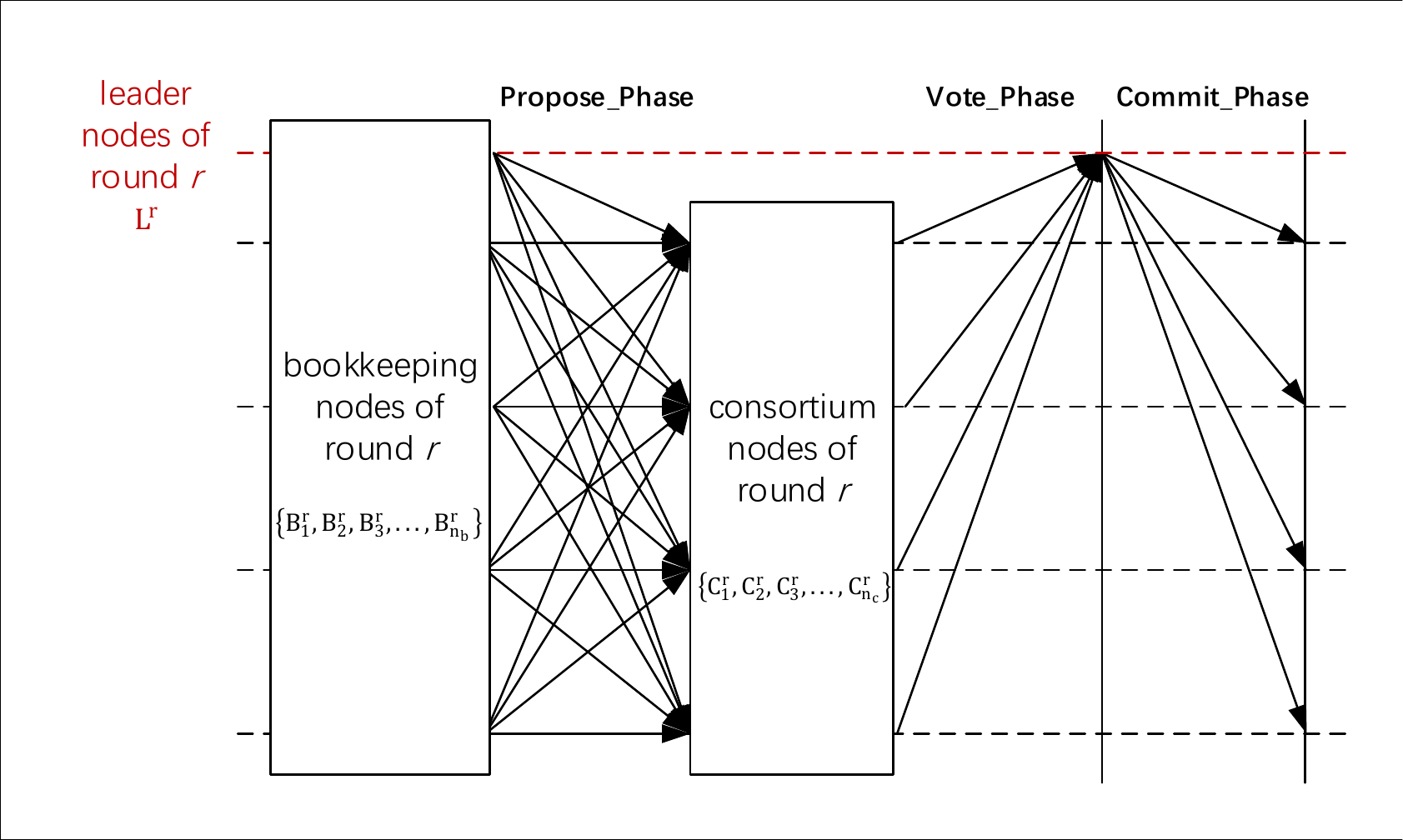}}
	\caption{Consensus Process of APoV}
	\label{fig2}
\end{figure}

The consensus process is shown in Figure.~\ref{fig2}.

Each round of the APoV consensus consists of the following steps.

S1: Each bookkeeping node generates a block and publishes it to the network. Each blockchain node collects all the blocks in this step.

S2: When the consortium node collects all the block generated in S1, it votes for each block and sends a total voting message to the leader node. The voting message contains the hash value of each block, as well as the agreed opinion and signature.

S3: The leader node collects the voting messages sent in S2 and counts the voting results. Statistical results and all voting messages will be stored in the block group header when the approval or disapproval of each block is more than half of the number of consortium nodes. The leader node then generates a random number as the number of the next leader node and writes it into the block group header. Finally, the leader node publishes the block group header to the network.

S4: When the blockchain node receives the block generated by the bookkeeping nodes and the block group header generated by the leader node, it will store them in the database as a block group.

\subsection{Improved FIB}

There are multiple identifiers including identity, content, geographic information, and IP address on the data plane of MIN. However, in the traditional FIB, the length of content names leads to a too large table and reduces lookup speed. On the one hand, we design the inter-translating algorithm with multiple identifiers. On the other hand, by building hash tables and prefix trees, we optimize the FIB algorithm to improve the lookup speed at massive scale.

\begin{algorithm}[!htbp]
	\caption{Insertion Algorithm}
	\label{Alg1}
	\raggedright
	\footnotesize
	{\bf Input:}\\
	\hspace*{0.1in}$H$: HT and trie-based FIB\\
	\hspace*{0.1in}$n$: $n=``/c_1/c_2/.../c_N"$ is the content name to insert\\
	\hspace*{0.1in}$f$: The corresponding forwarding information of $n$\\
	{\bf Output:}\\
	\hspace*{0.1in}$H$: HT and trie-based FIB, with $n$ inserted
	\begin{algorithmic}[1]
		\State lookup $n$ in HT
		\If{$n$ is the name of a real entry $(n,e)$}
		\State update $e$'s forwarding information with $f$
		\ElsIf{$n$ is the name of a non-real entry $(n,e)$}
		\State set $e$'s type to real, $e$'s forwarding information to $f$
		\For{each virtual entry $(\sim,e^*)$ in $e$'s subtree}
		\State set $e^*$'s type to semi-virtual
		\EndFor
		\Else
		\State create entry $(n,e_N)$ and insert it to HT
		\State set $e_N$'s type to real, $e_N$'s forwarding information to $f$
		\For{$i=N-1$ to 1}
		\State lookup $n_i=``/c_1/c_2/.../c_i"$ in HT
		\If{$n_i$ is the name of an entry $(n_i,e)$}
		\State add $e_{i+1}$ to $e$'s child list, set $e_{i+1}$'s parent to $e$
		\If{$e$ is virtual}
		\State set $e_j (i<j<N)$'s type to virtual
		\Else
		\State set $e_j (i<j<N)$'s type to semi-virtual
		\EndIf\\
		\hspace*{0.53in}\Return
		\Else
		\State create entry $(n_i,e_i)$ and insert it to HT
		\State add $e_{i+1}$ to $e_i$'s child list, set $e_{i+1}$'s parent to $e_i$
		\EndIf
		\EndFor
		\State add $e_i$ to $root$'s child list, set $e_i$'s parent to $root$
		\State set $e_j(0<j<N)$'s type to virtual
		\EndIf
	\end{algorithmic}
\end{algorithm}
\begin{algorithm}[!htbp]
	\caption{Deletion Algorithm}
	\label{Alg2}
	\raggedright
	\footnotesize
	{\bf Input:}\\
	\hspace*{0.1in}$H$: HT and trie-based FIB\\
	\hspace*{0.1in}$n$: $n=``/c_1/c_2/.../c_N"$ is the content name to delete\\
	{\bf Output:}\\
	\hspace*{0.1in}$H$: HT and trie-based FIB, with $n$ deleted
	\begin{algorithmic}[1]
		\State lookup $n$ in HT
		\If{$n$ is not the name of a real entry $(n,e)$}\\
		\hspace*{0.17in}\Return
		\EndIf
		\If{for $n$'s entry $(n,e)$, if $e$ is not a leaf}
		\State set $e$'s forwarding information to N/A
		\If{$e$'s parent is semi-virtual or real}
		\State set $e$'s type to semi-virtual
		\Else
		\State create an empty queue $q$ and insert $e$ into it
		\While{$q$ is not empty}
		\State $e^*=q$.pop()
		\State set $e^*$'s type to virtual
		\State insert all $e^*$'s semi-virtual child nodes into $q$
		\EndWhile
		\EndIf
		\Else
		\State remove $e$ from its parent's child list
		\State delete entry $(n,e)$ in HT
		\For{$i=N-1$ to 1}
		\State for $n_i=``/c_1/c_2/.../c_i"$ and its entry $(n_i,e_i)$
		\If{$e_i$ is non-real and $e_i$ is a leaf}
		\State remove $e_i$ from its parent's child list
		\State delete entry $(n_i,e_i)$ in HT
		\Else\\
		\hspace*{0.53in}\Return
		\EndIf
		\EndFor
		\EndIf
	\end{algorithmic}
\end{algorithm}

\begin{figure}[h]
\setlength{\abovecaptionskip}{0.cm}
\setlength{\belowcaptionskip}{-0.cm}
	\centerline{\includegraphics[width=3.3in]{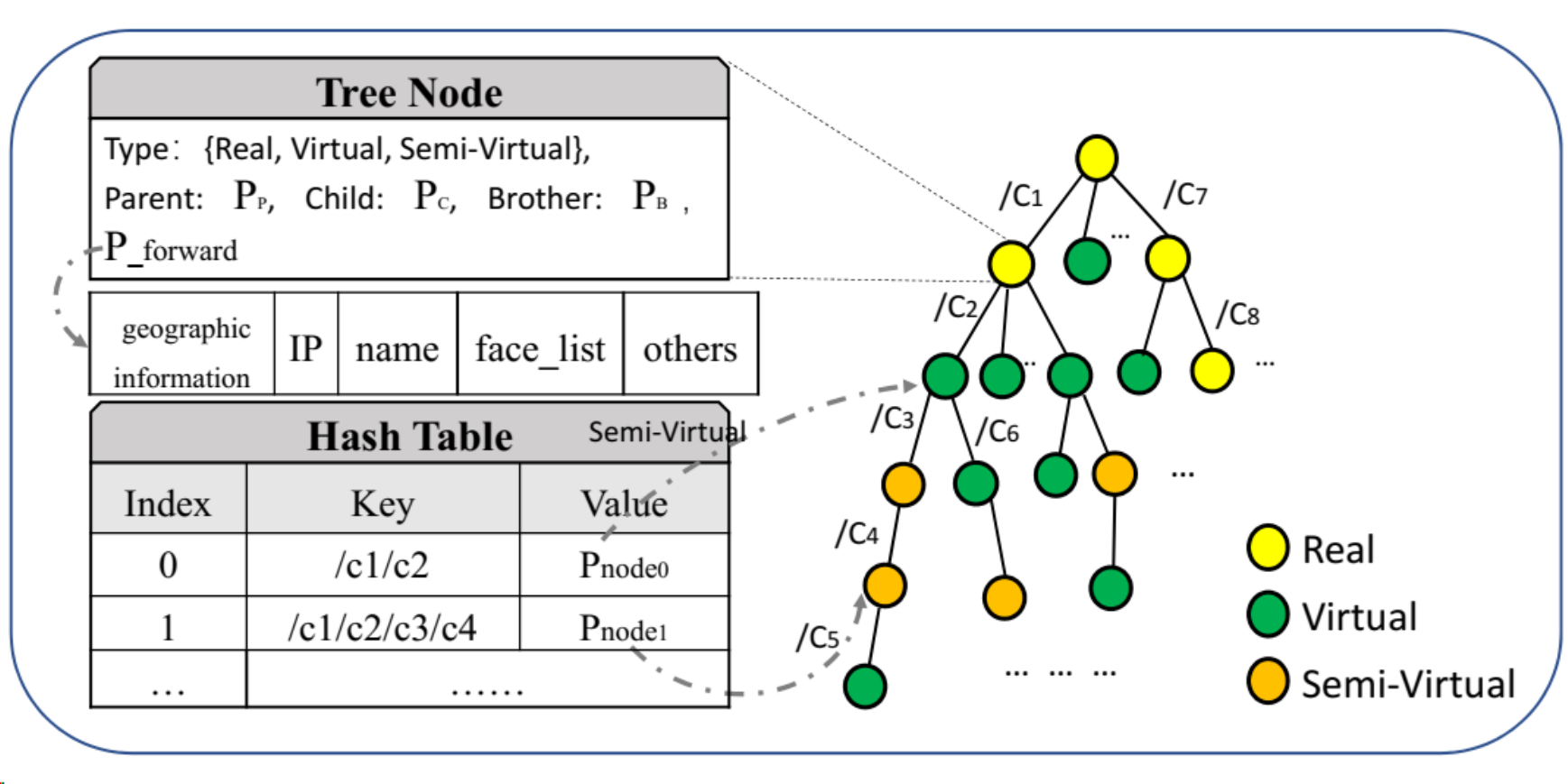}}
	\caption{The HPT-FIB Combining Hash Table and Prefix Tree}
	\label{fig3}
\end{figure}

~\\
\textbf{(1)Reconstructing HPT-FIB}

To support binary search, every true prefix of the content name stored in the table must also have corresponding entries. The process of checking whether prefixes exist and adding corresponding non-real entries is known as FIB reconstruction. Typically, in a reconstructed FIB, table entries can be divided into two categories: real entry and non-real entry. To avoid the false-negative error in traditional binary search, we subdivide non-real entry into the virtual entry and the semi-virtual entry.

The content names in the real entries referred to the actual data are used to forward Interest packets. All of the table entries are real before FIB reconstruction. Content names in non-real entries do not refer to data or guide the forwarding of Interest packets but are only used to support the binary search algorithm. A non-real entry is called a virtual entry if it has no real prefix. If the non-real entry has a real prefix, the table entry is called semi-virtual entry and requires backtracking before the end. We present the semi-virtual entry to avoid the false-negative error in HPT-FIB when the binary search process ends with a virtual entry.

The HPT-FIB is shown in Figure.\ref{fig3}. In the hash table, each content name (such as $/c1/c4/c5$) is calculated as the key, and its value points to the node in the prefix tree. Edges in the prefix tree represent a content name component (such as $/c1$). Each node represents a content name, which splices all the components on the path from this node to the root. In a tree node, there are five pieces of information: {state, parent node pointer, child node pointer, brother node pointer, forward pointer}, which maintain the structure of the tree.

Establishing HPT-FIB includes two basic operations: insertion and deletion, whose algorithms are shown in Algorithm.~\ref{Alg1} and Algorithm.~\ref{Alg2} respectively.

~\\
\textbf{(2)The Inter-translating Scheme}

In the network, routers in each domain maintain the HPT-FIB of multiple identifiers. This table records various identifiers of resources. The identifier inter-translating scheme is used to provide translation between different identifiers when users query resources.

When users register and publish content resources on MIN, the bounded identifiers are stored in the prefix tree whose array header is maintained by forwarding pointer ($ P_{\_Forward} $). When users query with content name, it routes without translation. When users query with another identifier, the network searches the $ P_{\_Forward} $ of the content, then selects the corresponding content name for routing.

~\\
\textbf{(3)The Lookup Algorithm}

In order to meet MIN's requirement of large-scale and fast routing, we design a high-speed lookup algorithm of HPT-FIB. It adopts binary search under the longest prefix matching (LPM) principle. As introducing semi-virtual entries, the return values for different HITs are different from traditional binary searches.
Specifically, there are three patterns based on the category of the last HIT entry.

\begin{itemize}
	\item If it is a real entry, the search for LPM is successful, and return the corresponding information.
	\item If it is a virtual entry, it is sure that there is no matching real prefix in the table, so return with no match.
	\item If it is a semi-virtual entry, there is at least one matching real prefix in the table. We can backtrack in the prefix tree to find the matching real prefix out and return it. Since backtracking in the prefix tree does not involve searching, this process has a minimal time overhead.
\end{itemize}

So, there are two kinds of lookup results. One is HIT, which means that there is a corresponding real entry in HPT-FIB (i.e., the last HIT entry is real entry or semi-virtual entry). The other is MISS, which means that the corresponding real entry does not exist (i.e., the last HIT entry is virtual). We test these two application scenarios in Section 4.2.

\subsection{Tunnel Transmission}

To promote the progressive deployment of MIN, we propose a tunnel scheme based on the new typed networks. Taking CCN as an example, this section describes the process of using it as a tunnel to transport IP packets.

~\\
\textbf{(1)Problem Description}

CCN \cite{K11} has revolutionized the address-centric transport architecture based on TCP/IP protocol. To realize the progressive deployment, TCP and CCN need to communicate mutually. There are many difficulties with this process. First, TCP is an end-to-end protocol that communicates through IP addresses and port numbers, which contradicts CCN's content-based philosophy. Second, in CCN, communication is a user-initiated process of ``pulling" the required data. However, in TCP, it is a ``push" process in which the sender sends data, and the receiver replies the acknowledgment message. The two are fundamentally different in semantics. Third, TCP ensures reliable end-to-end transmission, which CCN does not address.

~\\
\textbf{(2)Tunnel Scheme}

\begin{figure}[h]
\setlength{\abovecaptionskip}{0.cm}
\setlength{\belowcaptionskip}{-0.cm}
	\centerline{\includegraphics[width=3.3in]{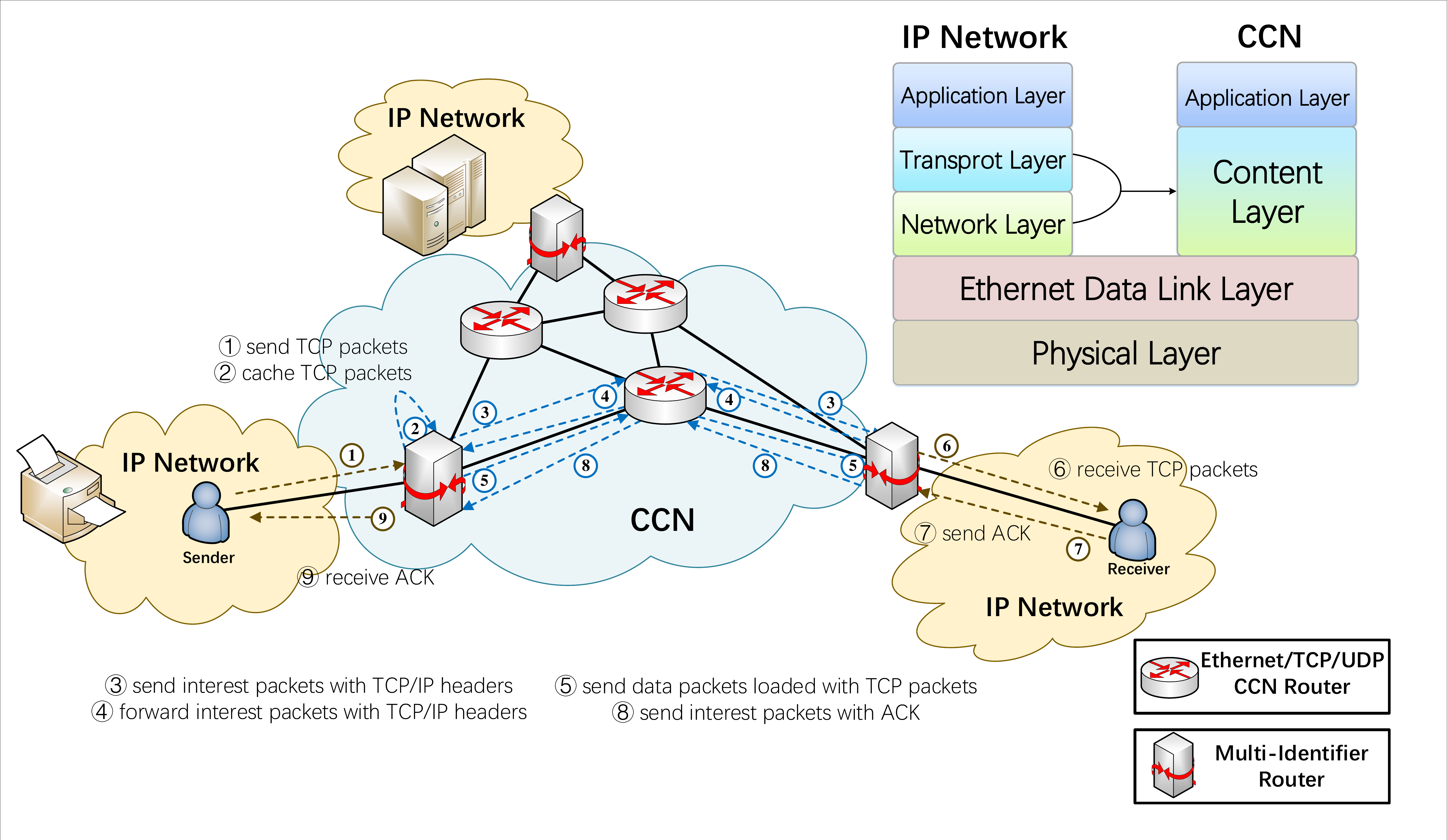}}
	\caption{The Schematic Diagram of IP-CCN-IP Transmission}
	\label{fig5}
\end{figure}

In the scheme, CCN is directly deployed on the MAC layer of Ethernet to get rid of the dependence on IP completely. The architecture of combining Ethernet transmission, TCP transmission, and UDP transmission is shown in Figure.\ref{fig5}.

To enable the two TCP ends to communicate through CCN, the scheme sets a pair of conversion nodes at the boundary between the IP network and the CCN. The transformation between identifiers and encapsulation of packets are carried out in MIR. Each MIR has a name mapped to its IP address as the routing prefix on the CCN, allowing CCN packets to flow smoothly to the designated multi-identifier router for further processing.

\begin{figure}[h]
\setlength{\abovecaptionskip}{0.cm}
\setlength{\belowcaptionskip}{-0.cm}
	\centerline{\includegraphics[width=3.3in]{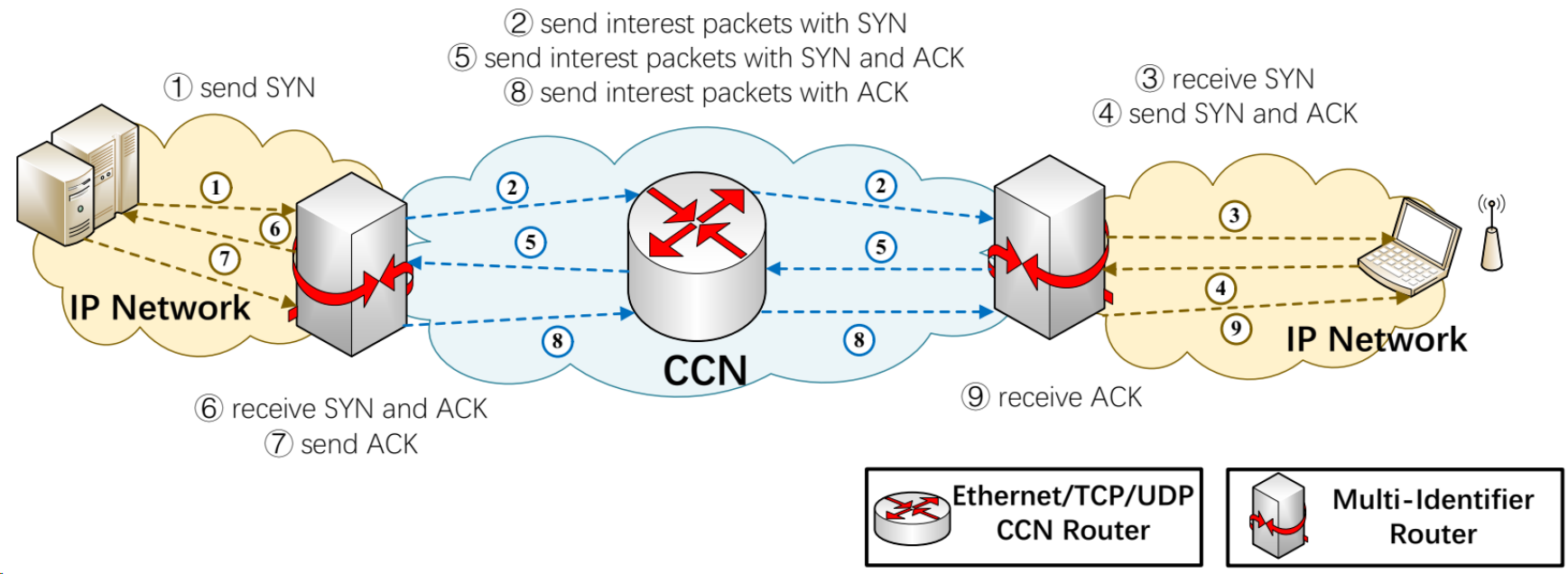}}
	\caption{Connection Establishment of IP-CCN-IP Transmission}
	\label{fig6}
\end{figure}

\textbf{Connection Establishment:} In this scheme, the three-way handshake to establish a connection between two TCP ends of IP networks is modified into the three-way Interest packet switches, as shown in Figure.\ref{fig6}. The TCP end sends SYN, SYN + ACK, and ACK control signaling to MIR. The CCN transports the TCP control signaling by encapsulating it into the Interest packet header. Finally, the receiving MIR forwards the control signaling to the other TCP end, thus completing the connection establishment of IP-CCN-IP transmission.

\textbf{Connection Termination: } The four-way handshake of connection termination between the two TCP ends is modified to the four-way Interest packet switches. The logic of Interest packets switching in this process is similar to that of the connection establishment.

\section{Prototype and experiment}

We developed a prototype of MIN and deployed it on the actual operators' network. This section tests and analyzes its ability on generation, management, and resolution, including the performances of the consensus algorithm, HPT-FIB query, and VoD service.

\subsection{Advanced Proof of Vote}

This section calculates the throughput of the blockchain consensus algorithm on the management plane of MIN. In APoV, since a new round of consensus can start only after the end of the current one, we calculate $ throughput $ through the time spent on each round of consensus. Consensus time consists of computation time and transmission time, that is,
\begin{equation}
  t_{cons}=t_{comp}+t_{tran}
  ~\\
\end{equation}

\textbf{(1)Calculation of the Transmission Time $ t_{tran} $}

According to the consensus steps in Section 3.1, in S1, the communication traffic of each bookkeeping node is the sum of block messages sent by it, and the communication traffic of each consortium node is all the block messages it receives. The communication pressure of the bookkeeping node is higher than that of the consortium node, so the transmission time in S1 is the communication time of the bookkeeping node,
\begin{equation}
  t_{tran}^1=(n_b+n_c-n_{bc}-1) \cdot (M+H+T \cdot K)/band
\end{equation}

where $ n_b $, $ n_c $ and $ n_{bc} $ are the number of bookkeeping nodes, consortium nodes and nodes concurrently holding these two identities respectively. $ M $, $ H $ and $ T $ are the size of the message header, the block header, and the transaction, respectively. $ K $ is the maximum number of transactions that can be placed within each block. $ band $ is the bandwidth of each node (assuming the same uplink and downlink bandwidth).

To balance the computing power between nodes, we assume that the leader node does not concurrently serve as the consortium node. In this case, the transmission time in S2 is
\begin{equation}
  t_{tran}^2=n_c \cdot (M+H_v+n_b \cdot V_b )/band
\end{equation}

where $ H_v $ and $ V_b $ are the size of the vote header and the single vote, respectively.

Similarly, the transmission time in S3 is
\begin{small}
\begin{equation}
  t_{tran}^3=(n_b+n_c-n_{bc}-1)[M+H_r+n_bR_b+n_c(H_v+n_b \cdot V_b)]/band
\end{equation}
\end{small}

where $ H_r $ and $ R_b $ are respectively the size of the voting result header and the voting result of a single block.

According to Equation (2)-(4), the transmission time is
\begin{equation}
  t_{tran}=t_{tran}^1+t_{tran}^2+t_{tran}^3
  ~\\
\end{equation}

\textbf{(2)Calculation of the Computation Time $ t_{comp} $}

Consider a simple network scenario of consortium blockchain with two servers, where Server A runs a blockchain node, and Server B runs multiple blockchain nodes. To reduce the waste of computing power, we set each node as both bookkeeping node and consortium node. We make the node on Server A the leader node.

Since the bandwidth in Server B is much larger than that between A and B, the transmission time of nodes on Server B can be regarded as 0, and the transmission time of Server A still follows the conclusion above. The advantage of this scenario is that it eliminates the impact of asynchronous transmission on performance in distributed networks, and only analyzes computational factors.

The blockchain parameters of the prototype are $ K=10000 $, $ M=266Byte $, $ H=692Byte $, $ T=40Byte $, $ H_v=400Byte $, $ V_b=100Byte $, $ H_r=170Byte $, $ R_b=400Byte $, $ band=1Gbps=125MB/s $, and the CPU of the server is Intel Xeon Silver 4114@2.20GHz. From the perspective of Server A, the states of the single blockchain node and the whole blockchain network can be observed simultaneously. We run 10 rounds of consensus at different scales. The time consumption of each step and each round is measured on Server A and averaged, as shown in Table.~\ref{pov}.


\begin{table*}[!t]
		\begin{center}
			\caption{Average Time of the Node on Server A in 10 Rounds of Consensus}
			\label{pov}
			\begin{tabular}{|m{3cm}<{\centering}|m{1.5cm}<{\centering}|m{1.5cm}<{\centering}|m{1.5cm}<{\centering}|m{1.5cm}<{\centering}|m{2cm}<{\centering}|m{3cm}<{\centering}|}
                \hline
                \multirow{2}*{Number of Nodes n} & \multicolumn{4}{c}{Time Consumption (s)} & & \multirow{2}*{Throughput (tx/s)}\\
                \cline{2-6}
                 & S1 & S2 & S3 & S4 & {A Round of Consensus}\tiny & \\
                 \hline
                 3 & 0.0311 & 0.0642 & 0.0255 & 0.0217 & 0.132 & 223706\\
                 \hline
                 4 & 0.0326 & 0.0750 & 0.0323 & 0.0268 & 0.150 & 263583\\
                 \hline
                 5 & 0.0377 & 0.0861 & 0.0295 & 0.0319 & 0.163 & 302719\\
                 \hline
                 6 & 0.0416 & 0.0986 & 0.0367 & 0.0377 & 0.189 & 315861\\
                 \hline
                 7 & 0.0470 & 0.113 & 0.0392 & 0.0419 & 0.217 & 322992\\
                 \hline
                 8 & 0.0505 & 0.130 & 0.0552 & 0.0477 & 0.252 & 314743\\
				\hline
			\end{tabular}
		\end{center}
	\end{table*}

According to the consensus steps in Section 3.1, the relationship between the time consumption of S1, S2 and S4 and the number of nodes $ n $ is linear. In S3, when $ n $ increases, the number of blocks that each consortium node needs to vote increases, so its time consumption can be described by a quadratic function. The time consumption of each step in Table.~\ref{pov} is fitted to
\begin{equation}
\begin{split}
  &t_{comp}^1=0.0041n+0.0174\\
  &t_{comp}^2=0.0130n+0.0229\\
  &t_{comp}^3=0.0012n^2-0.0082n+0.0415\\
  &t_{comp}^4=0.0052n+0.0062
\end{split}
\end{equation}

Further understanding of Equation (6) is that, $ t_{comp}^1 $ is reflected in the computation of the bookkeeping node, $ t_{comp}^2 $ in the computation of the consortium node, $ t_{comp}^3 $ in the computation of the leader node, and $ t_{comp}^4 $ in the computation of each node.

According to Equation (2)-(5), the transmission time $ t_{tran} $ is a cubic function of the number of nodes $ n $, so the consensus time $ t_{cons} $ can be described by a cubic function. The consensus time in Table.~\ref{pov} is fitted to
\begin{equation}
\begin{small}
t_{cons}=\dfrac{0.0312n^3-0.1920n^2+2.0714n+11.2500}{125}
\end{small}
\end{equation}

Substitute the parameter value of the prototype into Equation (2)-(5) to get the transmission time is
\begin{equation}
\begin{split}
t_{tran}&=t_{tran}^1+t_{tran}^2+t_{tran}^3\\
&=\dfrac{0.0001n^3+0.0008n^2+0.3213n-0.3214}{125}
\end{split}
\end{equation}

According to Equation (1), the computation time is
\begin{equation}
\begin{split}
t_{comp}&=t_{cons}-t_{tran}\\
&=\dfrac{0.0311n^3-0.1928n^2+1.7501n+11.5714}{125}
\end{split}
\end{equation}

Consider the further understanding of Equation (6). Take the computing power of the servers in the prototype as the standard. For a blockchain network with computing power $ a $, that is, the maximum computing power used by all nodes for consensus is $ a $ times of the standard computing power, then the minimum computation time is shown in Equation (10).

\newcounter{mytempeqncnt1}
\begin{scriptsize}
\begin{figure*}[htb]
\setcounter{mytempeqncnt1}{\value{equation}}
\setcounter{equation}{9} 
\begin{equation}
\begin{split}
t_{comp}^\prime=\frac{t_{comp}}{a}\cdot[1+\frac{(0.0012n^2+0.0141n+0.0880)}{n(0.0223n+0.0465)}]=\frac{(0.0235n^2+0.0606n+0.0880)(0.0311n^3-0.1928n^2+1.7501n+11.5714)}{a(2.7875n^2+5.8125n)}
\end{split}
\end{equation}
\setcounter{equation}{\value{mytempeqncnt1}}
\end{figure*}
\end{scriptsize}

According to Equation (5) and (10), the minimum consensus time in the network is
\setcounter{equation}{10}
\begin{equation}
t_{cons}^\prime=t_{comp}^\prime+t_{tran}
\end{equation}

To sum up, the upper limit of throughput within the blockchain network is shown in Equation (12).

\newcounter{mytempeqncnt2}
\begin{scriptsize}
\begin{figure*}[htb]
\setcounter{mytempeqncnt2}{\value{equation}}
\setcounter{equation}{11} 
\begin{equation}
\begin{split}
throughput=&k\cdot n/t_{cons}^\prime\\
=&10000n/[\frac{(0.0235n^2+0.0606n+0.0880)(0.0311n^3-0.1928n^2+1.7501n+11.5714)}{a(2.7875n^2+5.8125n)}+\frac{(0.0001n^3+0.0008n^2+0.3213n-0.3214)}{band}]
\end{split}
\end{equation}
\setcounter{equation}{\value{mytempeqncnt2}}
\hrulefill 
\end{figure*}
\end{scriptsize}

\begin{figure*}[htb]
\setlength{\abovecaptionskip}{0.cm}
\setlength{\belowcaptionskip}{-0.cm}
	\centerline{\includegraphics[width=8.6in]{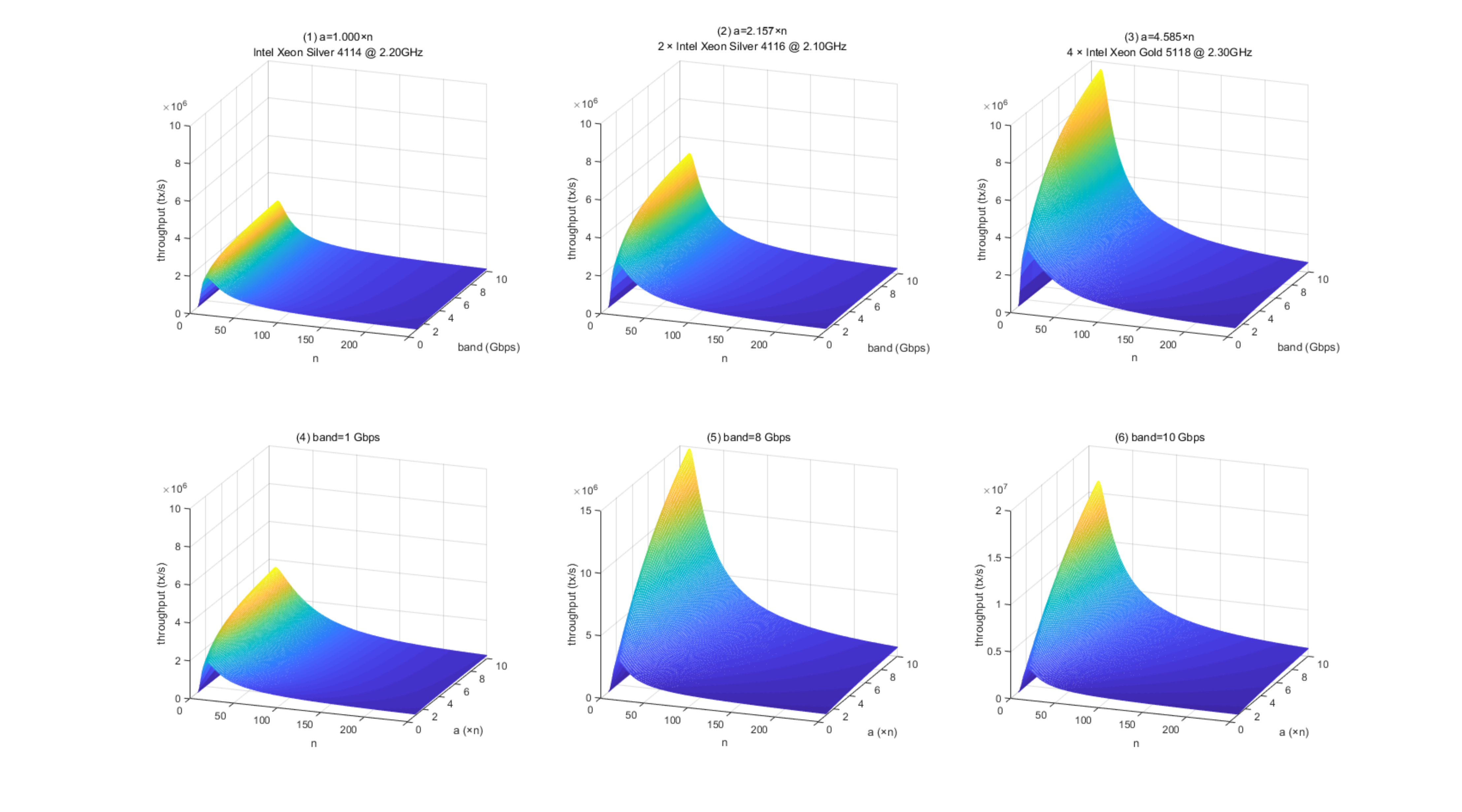}}
	\caption{The Influence of Node Number $ n $, Computing Power $ a $ and Bandwidth $ band $ on the Upper Limit of Throughput in the APoV Blockchain}
	\label{fig8}
\end{figure*}

Based on Equation (12), it is possible to estimate the upper limit of performance in the real blockchain network composed of servers and switches using APoV consensus algorithm. In the consortium blockchain network, each server typically runs only one node. Assuming that the other configurations of nodes are the same, when their CPUs are Intel Xeon Silver 4114@2.20GHz, Intel Xeon Silver 4116@2.10GHz, and Intel Xeon Gold 5118@2.30GHz respectively\footnote{Data Sources: http://cdn.malu.me/cpu/}, the upper limit of throughput is affected by $ n $ and $ band $, as shown in Figure.\ref{fig8} (1)-(3). When the bandwidth of nodes is set as 1Gbps, 8Gbps and 10Gbps respectively, the influence of $ n $ and $ a $ on the upper limit of throughput is shown in Figure.\ref{fig8} (4)-(6).

When the number of nodes is small (generally less than 10), the computing power used for consensus is not fully utilized, so the number of nodes is the main factor affecting the throughput. When the number of nodes increases, the performance can be approximately increased with the improvement of computing power and bandwidth.

At present, the throughput of most blockchain consensus algorithms is less than 100 thousand transactions per second\cite{W1}. We measure the prototype consisting of 20 blockchain nodes, each of which is responsible for both voting and bookkeeping. The APoV consensus can achieve a stable throughput of more than 300 thousand transactions per second. Although the actual value is much lower than the theoretical value in Figure.~\ref{fig8}, it greatly exceeds other consensus algorithms. We will continue optimizing the algorithm code of APoV to utilize the computing power and bandwidth in the network entirely.

\subsection{FIB Combining Hash Table and Prefix Tree}

This section calculates the scale and throughput of HPT-FIB algorithm on the data plane of MIN.

~\\
\textbf{(1)Scalability}

The data source of the experiment is the self-generated CCN content names. As CCN has not deployed on a large scale, it is difficult to obtain large-scale CCN dataset from the real network. Therefore, we extract statistical features from the current network flow and obtain a large number of random URL-like names through simulation to generate a large-scale HPT-FIB. The test of building time for HPT-FIB is shown in Table.~\ref{FIB-G}.

\begin{footnotesize}
\begin{table}[!t]
\setlength{\abovecaptionskip}{0.cm}
\setlength{\belowcaptionskip}{-0.cm}
	\begin{center}
		\caption{The Time for Generating HPT-FIB Entries}
		\label{FIB-G}
		\begin{tabular}{|m{1.2cm}<{\centering}|m{0.6cm}<{\centering}|m{0.8cm}<{\centering}|m{0.8cm}<{\centering}|m{0.8cm}<{\centering}|m{0.8cm}<{\centering}|m{0.8cm}<{\centering}|}
			\hline
			{Size of FIB (billion)} \tiny &0.1 &1.0 &2.0 &2.5 &3.0 &3.5\\
			\hline
			Run Time(s) & 187.58 & 1649.75 &3723.98 &4925.64 &6271.49 &7760.69\\
			\hline
			{Actual Size of FIB (billion)} \tiny &0.1 &1.0 &2.0 &2.5 &3.0 &3.5\\
			\hline
		\end{tabular}
	\end{center}
\end{table}
\end{footnotesize}

The test result indicates that the HPT-FIB in this scheme can support large-scale prefix of content names storage.

~\\
\textbf{(2)Lookup Performance}

In this section, we test the lookup speed of HPT-FIB. The contrast algorithm includes LPM-based linear search and LPM-based binary search without backtracking in the prefix tree. As mentioned above, there are two kinds of lookup results: HIT and MISS. We test these two application scenarios.

For every testing, we lookup 500,000 times in HPT-FIB, which contains 5 million real entries. The above experiments are carried out ten times, and we count up their average values for evaluation and analysis. For the sake of comparison, the measurement is set to Operation Throughput, which is inversely proportional to the average operating time. Also, the throughput rate for linear search is set to 1 as a benchmark.

When the average length of the content names to be searched is $N$, and the result is all MISS, which means that none inquired content name correspond to a real entry in HPT-FIB, the performances are shown in Table.~\ref{FIB-SM}.

\begin{table}[!t]
\setlength{\abovecaptionskip}{0.cm}
\setlength{\belowcaptionskip}{-0.cm}
	\begin{center}
		\caption{The Throughput of Name Search with all MISS}
		\label{FIB-SM}
		\scriptsize
		\begin{tabular}{|m{0.2cm}|m{1.2cm}<{\centering}|m{1.2cm}<{\centering}|m{1.2cm}<{\centering}|m{1.2cm}<{\centering}|m{1.2cm}<{\centering}|}
			\hline
			\multirow{2}*{$N$} & \multicolumn{2}{|c|}{Linear Search} & \multicolumn{3}{|c|}{Binary Search}\\
			\cline{2-6}
			& Average Searching Time &Throughput&Average Searching Time& Throughput (No Backtrack) & Throughput (HPT-FIB)\\
			\hline
			6 & 6 & 100\% & 2.34 & 236\% & 235\%\\
			\hline
			7 & 7 & 100\% & 2.54 & 250\% & 247\% \\
			\hline
			8 & 8 & 100\% & 2.70 & 267\% & 266\% \\
			\hline
			9 & 9 & 100\% & 2.85 & 280\% & 279\% \\
			\hline
			10 & 10 & 100\% & 2.96 & 301\% & 299\% \\
			\hline
		\end{tabular}
	\end{center}
\end{table}

When lookup does not HIT, the linear search needs to traverse all prefixes of content names. The time cost is proportional to the length of content names, causing problems about efficiency and security. In contrast, binary search reduces the computational overhead of the LPM algorithm to the logarithmic level, which significantly improves the throughput. Also, since our algorithm needs no backtracking when lookup ends without HIT, the time cost is close to the original binary search. As a result, we solve the false negative error with acceptable costs.

For the lookup test of all HIT which means that every inquired content name corresponds to a real entry in HPT-FIB, we assume that the average length of content names is $M$ with distribution $ \rho(N;\bar{\lambda}) $. Since the average length of domain names is 2 or 3 and CCN content names are generally longer than URL requests in HTTP, we use $M = 4,5$ as the test parameter. The performances of all HIT are shown in Table.~\ref{FIB-SH}.

\begin{table}[!t]
\setlength{\abovecaptionskip}{0.cm}
\setlength{\belowcaptionskip}{-0.cm}
	\begin{center}
		\caption{The Throughput of Name Search with all HIT}
		\label{FIB-SH}
		\scriptsize
		\begin{tabular}{|m{0.2cm}|m{0.2cm}|m{1.2cm}<{\centering}|m{1.0cm}<{\centering}|m{1.2cm}<{\centering}|m{1.2cm}<{\centering}|m{1.0cm}<{\centering}|}
			\hline
			\multirow{2}{*}{M} & \multirow{2}*{N} & \multicolumn{2}{|c|}{Linear Search} & \multicolumn{3}{|c|}{Binary Search}\\
			\cline{3-7}
			& & Average Searching Time & Through-put & Average Searching Time & Through-put (No Backtrack) & Through-put (HPT-FIB)\\
			\hline
			\multirow{5}*{3}
			& 6 & 3.98 & $ 100\% $ & 2.84 &$  126\% $ & $ 123\% $\\
			\cline{2-7}
			&7 & 4.99 & 100\% & 3.02 & 131\% & 130\% \\
			\cline{2-7}
			&8 & 6.02 & 100\% & 3.17 & 151\% & 150\% \\
			\cline{2-7}
			&9 & 7.00 & 100\% & 3.31 & 168\% & 167\% \\
			\cline{2-7}
			&10 & 8.01 & 100\% & 3.45 & 185\% & 183\% \\
			\hline
			\multirow{5}*{4}
			&6 & 3.03 & 100\% & 2.90 & 85\% & 85\%\\
			\cline{2-7}
			&7 & 4.00 & 100\% & 3.06 & 104\% & 101\% \\
			\cline{2-7}
			&8 & 5.02 & 100\% & 3.21 & 123\% & 124\% \\
			\cline{2-7}
			&9 & 6.02 & 100\% & 3.35 & 141\% & 139\% \\
			\cline{2-7}
			&10 & 7.03 & 100\% & 3.47 & 159\% & 159\% \\
			\hline
		\end{tabular}
	\end{center}
\end{table}

When lookup ends with HIT, the average time cost of the linear search is proportional to $(N-M+ 1)$. The time of binary search is almost not affected by $M$ and is approximately proportional to $log(N)$. For the scenarios with long content names to be searched and short table entries, the HPT-FIB lookup is faster.

According to the test results, the backtracking does not affect lookup speed because of no string searching. Therefore, the proposed algorithm solves the false negative error in random search with minimal efficiency loss.

\subsection{Tunnel Transmission}

\begin{figure*}[!t]
\setlength{\abovecaptionskip}{0.cm}
\setlength{\belowcaptionskip}{-0.cm}
	\centerline{\includegraphics[width=7.0in]{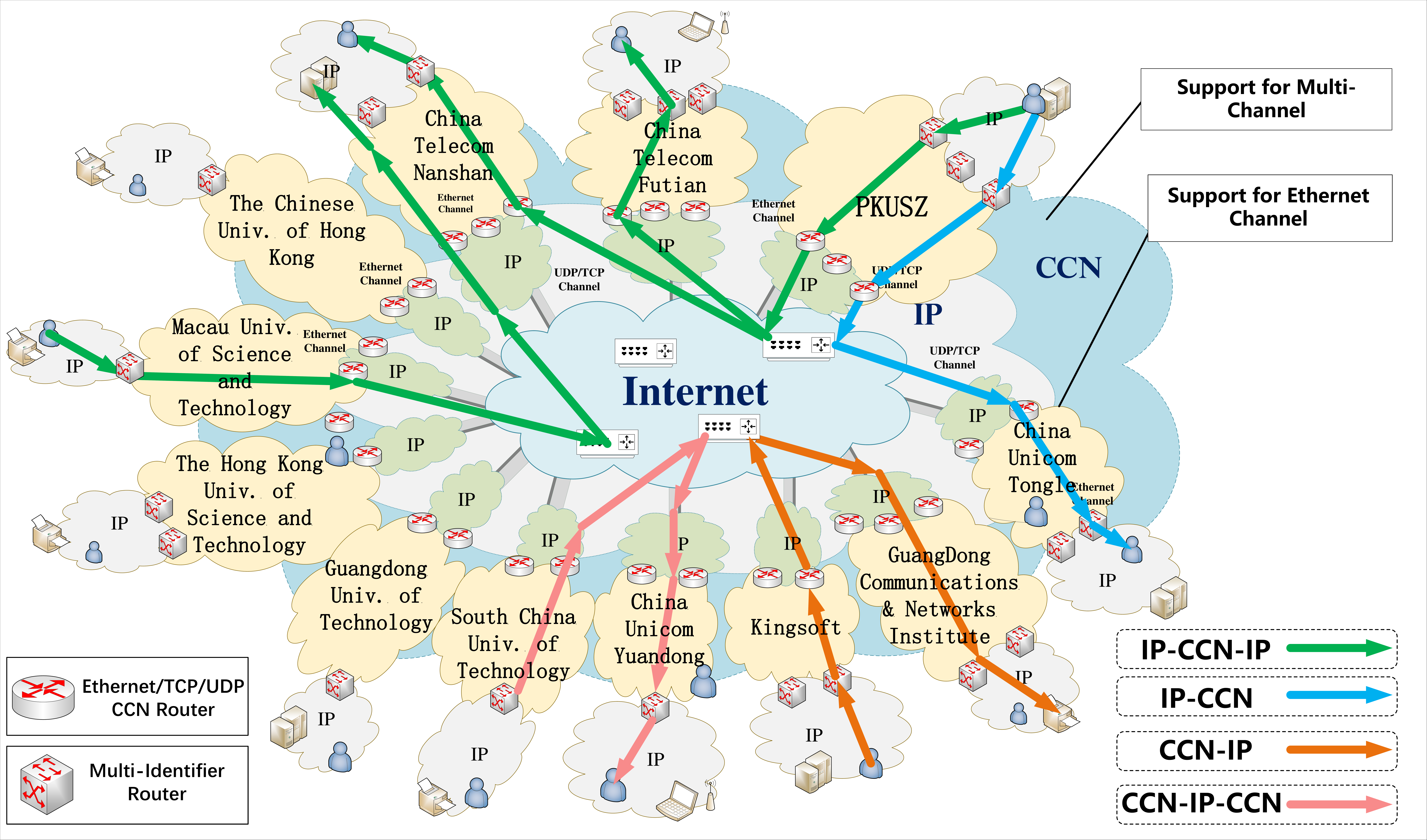}}
	\caption{Topology of the Prototype}
	\label{fig12}
\end{figure*}

The prototype realizes tunnel transmission under various network scenarios, including IP-CCN-IP, IP-CCN, CCN-IP, and CCN-IP-CCN. The topology is shown in Figure.\ref{fig12}. Based on this, the prototype can provide High Definition Video on Demand (HDVoD) service with multiple transmission channels.

The exporting bandwidth of the servers in Peking University Shenzhen Graduate School (PKUSZ) is 100M special broadband, and that of other institutions is also more than 50M. On top of the bandwidth constraint, the network environment differences between operators and between mainland China, Hong Kong, and Macao may limit the transmission rates. The following are the experimental results of our prototype under four scenarios.

~\\
\textbf{(1)IP-CCN-IP Transmission}

\begin{figure}[h]
	\setlength{\abovecaptionskip}{0.cm}
	\setlength{\belowcaptionskip}{-0.cm}
	\centerline{\includegraphics[width=3.3in]{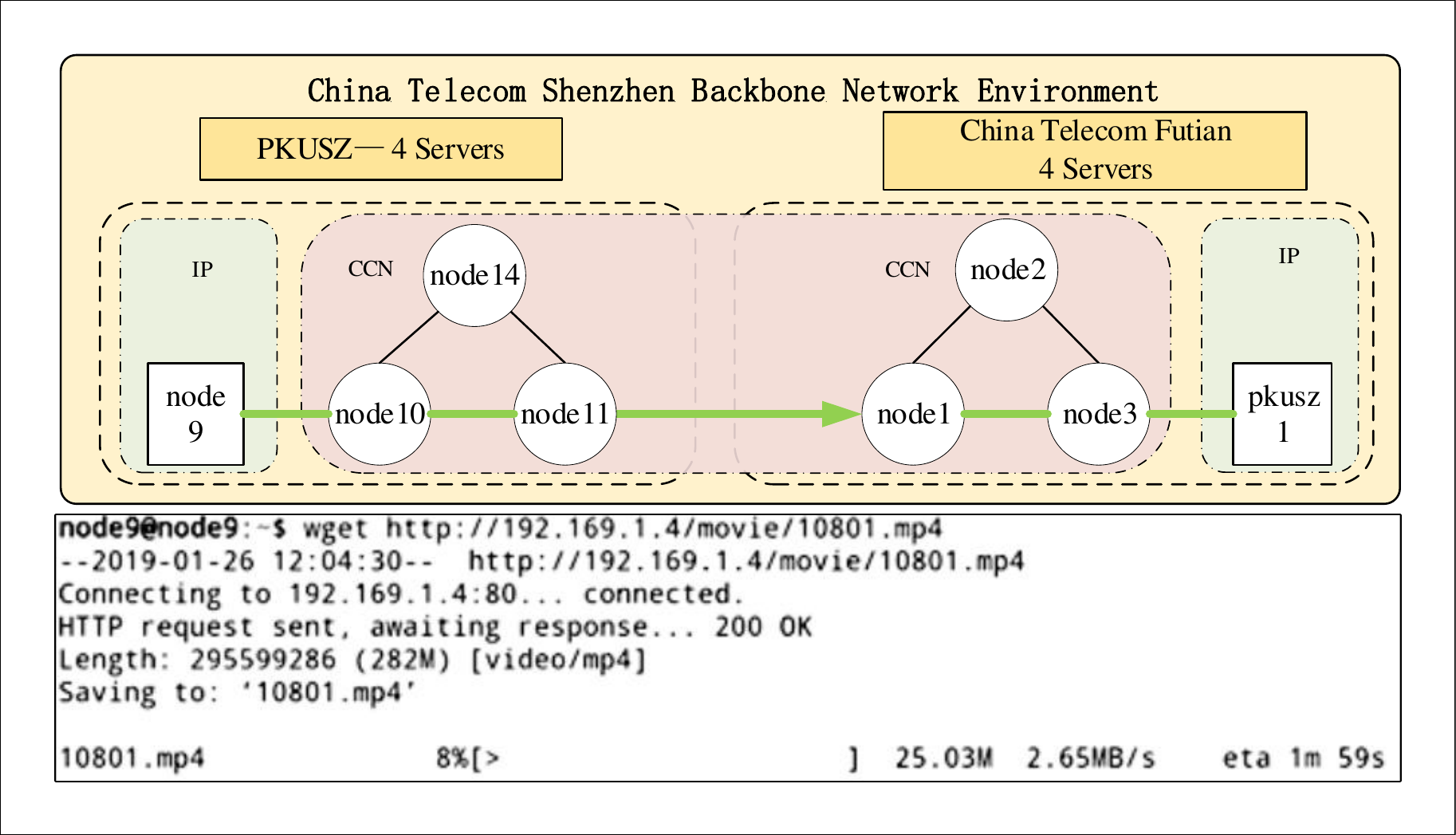}}
	\caption{The Detailed Topology and the Experimental Result (Pull from $ pkusz1 $ to $ node9 $)}
	\label{fig13}
\end{figure}

\begin{figure}[h]
	\setlength{\abovecaptionskip}{0.cm}
	\setlength{\belowcaptionskip}{-0.cm}
	\centerline{\includegraphics[width=3.3in]{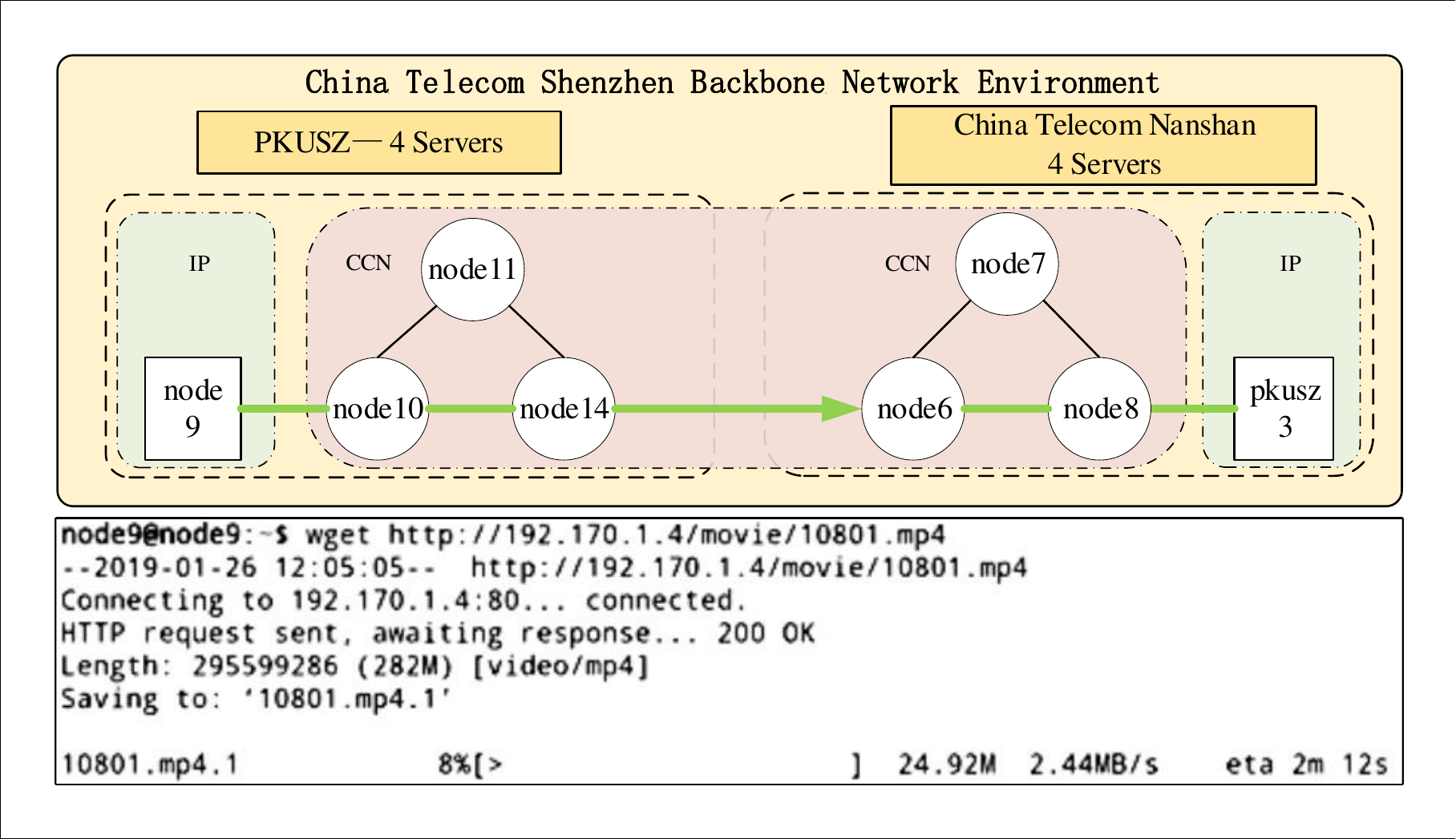}}
	\caption{The Detailed Topology and the Experimental Result (Pull from $ pkusz3 $ to $ node9 $)}
	\label{fig14}
\end{figure}

Video resources are uploaded on the node $ pkusz1 $ of China Telecom Futian and the node $ pkusz3 $ of China Telecom Nanshan. We first use $ node10 $ of PKUSZ as MIR and $ node9 $ of PKUSZ as the IP node to pull the video from $ pkusz1 $ to $ node9 $. Figure.\ref{fig13} shows the detailed topology and the experimental result. It can be seen that the transmission rate reaches 2.65MB/s.

We then pull the video resource from $ pkusz3 $ to $ node9 $. The detailed topology and the experimental result are shown in Figure.\ref{fig14}, where the transmission rate is 2.44MB/s.

\begin{figure}[h]
\setlength{\abovecaptionskip}{0.cm}
\setlength{\belowcaptionskip}{-0.cm}
	\centerline{\includegraphics[width=3.3in]{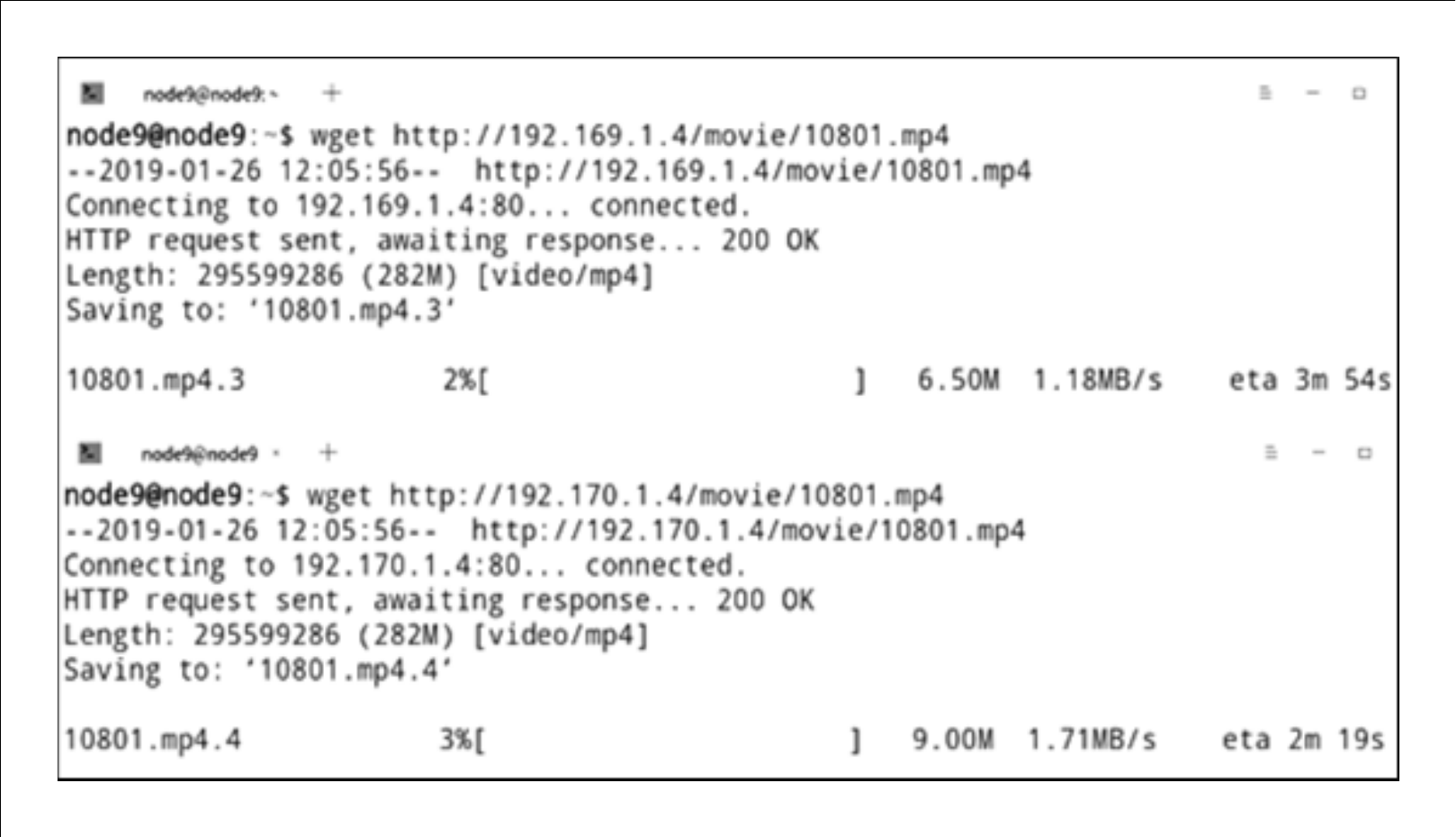}}
	\caption{The Experimental Result (Pull to $ node9 $ from $ pkusz1 $ and $ pkusz3 $ Simultaneously)}
	\label{fig15}
\end{figure}

When video resources on $ pkusz1 $ and $ pkusz3 $ are pulled to $ node9 $ simultaneously, the rates of the two transmission channels are 1.18MB/s and 1.71MB/s respectively, as shown in Figure.\ref{fig15}. The transmission rate is lower than before when pulling from a single node, but the total transmission rate remains the same. The main reason is that the access bandwidth of $ node9 $ is limited, resulting in the performance bottleneck.

~\\
\textbf{(2)IP-CCN Transmission}

The video resource is uploaded on the node $ host2 $ of China Unicom Tongle in CCN. As with the IP-CCN-IP transmission, we get the video from $ host2 $ to $node9$ through $node10$. The tested transmission rate is 0.65MB/s.

~\\
\textbf{(3)CCN-IP Transmission}

The video resource is uploaded on the node $gdcni1$ of Guangdong Communications \& Networks Institute (DGCNI) in the IP network. Similarly, we pull the video from $gdcni1$ to the node $host1$ of Kingsoft. The tested transmission rate is 1MB/s.

~\\
\textbf{(4)CCN-IP-CCN Transmission}

We have also implemented the CCN-IP-CCN transmission in the prototype of MIN. In this experiment, we place the video resource on the node $host2$ of China Unicom Yuandong, and pull it to the node $SCUT$ of South China University of Technology (SCUT). The tested transmission rate is 2.65MB/s.

\section{Conclusion and Future Work}

A future network should be decentralized, secure and compatible with the existing IP-based network. In this paper, we propose a Multi-Identifier Network that constructs a network layer with parallel coexistence of multiple identifiers, including identity, content, geographic information, and IP address. The network provides the generation, management, and resolution services of identifiers and use consortium blockchain to enable decentralized management. To further accelerate the forwarding process, we improve HPT for FIB by combining hash table and prefix tree. In addition, we propose a scheme of transporting IP packet using CCN as a tunnel to support progressive deployment. Finally, we develop the prototype to operators' networks and verify its performances of the consensus algorithm, FIB query, and VoD service. The test results illustrate that the network achieves excellent performance and can support real-world applications after further development.

\bibliographystyle{ACM-Reference-Format}
\bibliography{sample-base}

\appendix

\end{document}